# An Optimized Self-Adaptive Thermal Radiation Turn-Down Coating with Vanadium Dioxide Nanowire Array

Ken Araki & Richard Z. Zhang*

*Department of Mechanical Engineering, University of North Texas, Denton, 76207, USA.*

**ABSTRACT:** High performance metasurfaces for thermal radiative cooling applications can be identified using computational optimization methods. This work has identified an easy-to-fabricate temperature phase transition $VO_2$ nanowire array laid atop dielectric $BaF_2$ Fabry-Perot cavity-on-metal with total coating thickness of 2 μm. This optimized structure has ability to self-adaptively switch between high reflectance at low temperature to high emissivity at high temperature in the broad thermal infrared spectrum. This design demonstrates exceptional turn-down figure-of-merit compared to previously realized configurations utilizing $VO_2$ metasurfaces and multilayers. The mechanism is achieved with a sub-wavelength nanowire array effective medium that switches between anti-reflecting gradient coating and Fabry-Perot interference. This thin metasurface coating could impact self-cooling of the solar cells, batteries, and electrical devices where risk presents at high temperatures.

---

* Corresponding author: zihao.zhang@unt.edu (R. Z. Zhang).

**Nomenclature**

| | |
|---|---|
| $b$ | space between nanowires, μm |
| $d$ | thickness, μm |
| DCM | Daytime Cooling Merit |
| **E** | electric field vector |
| $f$ | phase transition filling ratio |
| FOM | Turn-Down Figure of Merit |
| $h$ | nanowire height, μm, |
| $\hbar$ | reduced Planck's constant [J·s] |
| $I$ | radiative intensity or solar irradiance, W/sr·m$^2$/μm |
| $k_B$ | Boltzmann constant, J/K |
| $n$ | refractive index |
| $N$ | number of samples |
| $P$ | radiative power, W/m$^2$ |
| PDD | power dissipation density, W/m$^3$ |
| **r** | randomization array |
| $T$ | temperature, K |
| $w$ | width of the nanowire, μm |
| $W$ | width of the temperature range of metal-insulator transition, K |

*Greek symbols*

| | |
|---|---|
| ε | emissivity |
| $\phi$ | filling ratio |
| $\varphi$ | azimuthal angle, deg |
| $\theta$ | incident angle, deg |
| $\kappa$ | extinction coefficient |
| Λ | period, μm |
| λ | wavelength, μm |
| $\omega$ | angular frequency, rad/s |
| $\rho$ | reflectivity |
| $\tau$ | transmissivity |

*Subscripts*

| | |
|---|---|
| 0 | vacuum / center value |
| AM1.5 | air mass coefficient 1.5 |
| BB | black body |
| coat | coating |
| con | combined convection and conduction heat transfer |
| cool | cooling |
| D | dielectric |
| EMT | effective medium theory |
| h | halfway through meta-insulator transition |
| i | insulator |
| m | metal |
| TE | transverse electric |
| TM | transverse magnetic |

## 1. INTRODUCTION

Optical properties of thin films can be manipulated by introducing multilayered nano-patterns or periodic nanoelements that interface with electromagnetic waves from the visible to infrared. Various flavors of engineered nanoscale coatings include sub-diffraction plasmonic gratings [1-3], ultra-high reflection high contrast gratings [4-6], traditional diffraction gratings [7], and simple multilayers such as Fabry-Perot quarter-wave layers [8-10]. Among the considered, these nanoscale metasurfaces are potentially useful in photovoltaic device improvement and thermal emission regulation [11-14]. Yet the key attribute to be sought after is thermochromic tuning – the ability to change optical properties with engineered surface temperature – which may result in self-cooling amid temperature or environmental changes. Temperature phase change materials, such as vanadium dioxide ($VO_2$) [9, 15, 16], $SmNiO_3$ [17, 18], $Ti_3O_5$ [19], and many others can tailor the optical properties when things get heated.

$VO_2$ integrated in photonic multilayers and gratings have demonstrated thermal radiative switching [9, 15, 20] A common coating design utilizes the concept of Fabry-Perot cavity interference to obtain low emittance when cool and high emittance at hot [9]. More complex multilayers can be added, for the purpose of daytime radiative cooling, as absorption to solar radiation is minimized compared to the emission in the infrared [15]. Other designs have patterned at-wavelength size scale cylindrical gratings [21, 22]. So far, the turn-down thermal performance of such coatings has been less than anticipated. The optimal thermochromic metastructure needs to be designed by weighting the Planck blackbody distribution at the corresponding wavelengths and temperatures close to the phase transition, as determined by Wien's displacement law. This work also considers the optimized structure that is easily feasible for nano/micro-fabrication, for a future in large scale-produced flexible radiative cooling blankets.

This work proposes a high turn-down grating by computational optimization of $VO_2$ nanowires in a thin film. $VO_2$ has a phase transition temperature at 340 K [14, 23, 24], where the phonon-electron structure transforms from insulating to metallic at temperatures above this transition temperature [25]. By randomized trials of at-and sub-wavelength deposited $VO_2$ nano/micro-features, a periodic $VO_2$ nanowire array and a solid $VO_2$ monolayer that sandwiches a quarter-wave thick lossless dielectric. Fabricated $VO_2$ nanowires typically of 50-150 nm in diameter were recently used demonstrating the breakdown on the Franz-Wiedemann law in thermoelectricity [25]. These $VO_2$ nanowires were fabricated through vapor transport method, a bottom-up method [25-28], and these $VO_2$ nanobeams can be laid and aligned in-plane, compared to the top-down photolithography and etching processes for traditional at-wavelength gratings [24]. This bottom-up method can greatly reduce complexity in fabrication, where minimal coating thickness is needed.

We utilize a Monte-Carlo optimization method to obtain the best nanowire array period and width of square cross-section $VO_2$ nanowires [29]. This method is applied to obtain the $VO_2$ monolayer as well, which serves as a gradual index anti-reflection layer above the fully-reflecting silver substrate. The broadband reflectance or broadband emissivity turn-down performance is quantified with a turn-down performance figure-of-merit (FOM), and compared with other $VO_2$ phase transition passive thermal radiation self-cooling coatings. The proposed structure has a total thickness of about 2.5 μm, including the 500 nm silver substrate. In this design, the coating on a flexible polymer backing sheet can be applied to alleviate the loss in efficiency for solar cells, whose temperatures can be as high as 80°C in summers under highly irradiative climates, such as Texas. The flexible coating can be applied to one or both sides of a vacuum gap between a solar cell housing and the roof. Similarly, it can be utilized for cooling compact processor components that gets as warm as 80 to 100°C. Other usages include thermal management for lithium-ion cells casings, passive-switched spacecraft insulation blankets, and other common Earth-temperature thermal regulation applications.

## 2. COMPUTATIONAL METHODS

### *2.1. Theory of Optical Coatings*

In this paper, Rigorous Coupled-Wave Analysis (RCWA), a well-established multilayered diffraction method to solve Maxwell equations [30, 31], is utilized to calculate the optical radiative properties of the optimized periodic coatings. The materials and geometrical parameters of the proposed coating is shown in Figure 1: The grating has a period of $\Lambda$, nanowire width of $w$, nanowire height of $h$, distance between nanowires $b$, thickness of the lossless dielectric is $d_D$, and thickness of the $VO_2$ sub-monolayer is $d_{VO_2}$. Initially, the nanowire aspect ratio can be non-square to demonstrate possible effects.

In designing toward desirable high reflection or high emission properties, selection of optical parameters is important in tailoring its optical parameters. The contrast of the dielectric layers helps construct an anti-reflecting coating, where each transparent layer satisfies the geometric mean of the surround layers, such that $n_1 = \sqrt{n_0 n_2}$ [32]. The underlying totally reflecting metal substrate can vary among aluminum, silver, gold, and others that are opaque and reasonably produced at sub-micron thickness. The anti-reflecting layer thicknesses must also satisfy quarter-wave law, such that $h = \lambda/4n$. This quarter-wave method is widely used to determine broadband reflectance at certain wavelength range known as Bragg reflector, where high- and low- refractive index materials are stacked into multilayers with quarter-wave thickness [33]. On the other hand, Fabry-Perot resonance are formed by constructive interference of electromagnetic waves within the dielectric spacer sandwiched by metallic thin films. Ideally, these metallic boundaries are infinitesimally-thin and lossless. This produces perfect absorptivity or emissivity, and perfect transmissivity at moded frequencies [8, 9]. The switching between one to the other mechanism is how our turn-down capability is designed. In addition, for periodic features at the sub-wavelength, such that $\Lambda << \lambda$, effective medium theory (EMT) can be applied to approximate the optical properties of gratings as expressed by a single layer composed of a filling ratio of dielectric functions [31, 34].

### *2.2. Radiative Properties*

The dielectric function of $VO_2$ are obtained from empirical spectroscopy of its thin films [35]. Its optical properties are expressed with effective medium and Fermi-Dirac like distribution as shown,

$$\varepsilon_{\mathrm{eff}}^{0.3} = (1-f)\,\varepsilon_{\mathrm{i}}^{0.3} + f\varepsilon_{\mathrm{m}}^{0.3} \tag{1}$$

$$f(T) = \frac{1}{1+\exp\left[\dfrac{W}{k_{\mathrm{B}}}\left(\dfrac{1}{T}-\dfrac{1}{T_{\mathrm{h}}}\right)\right]} \tag{2}$$

where $\varepsilon_i$ and $\varepsilon_m$ is the dielectric function for insulator and metal $VO_2$ respectively. $W$=3.37 K is width of the temperature range of metal-insulator transition. The partial monoclinic-to-rutile lattice phase transition occurs at $T_h$ = 78.5°C, where at this temperature, 50% of $VO_2$ is metallic [35]. Thus, the onset transition temperature of $VO_2$ from insulator to metal is 68°C. In Fig. 2 (a), the imaginary part of the refractive index in $VO_2$ for the cold state is close to zero in the mid- to far-infrared (around 4 to 10 μm). This infrared transparent region is highlighted in gray, and can play an important role in low absorption. On the other hand, it is absorbing in visible and far-infrared region.

As shown in Fig. 2(b), there are a few broadly infrared-transparent dielectric candidate materials that can be utilized as the dielectric spacer. These include barium fluoride ($BaF_2$), zinc sulfide (ZnS), and silicon (Si). The most common, silicon, has broadly constant refractive index of around $n = 3.4$, with no extinction coefficient provided undoped low number density and optical quality [31]. For $BaF_2$ and ZnS, the refractive index is lower, but have absorption component in far infrared region, due to the Reststrahlen effect of gaps in the upper-level electron bands [36, 37] The lower refractive index is appreciated in the arrangement of anti-reflection gradient index of layers. In certain other turn-down coating designs, $MgF_2$ can also be the candidate for the dielectric materials. However, it has a significant absorption near the Reststrahlen band, which may result in reduction in the performance of radiative properties [38]. Insulators with absorption in infrared wavelengths up to 30 μm must be avoided, as a wavelength window cutoff in the upper limit leaves more than 10% of Planck radiation at atmospheric temperatures unaccounted for. The balance of broad infrared transparency and close to unity refractive index needs to be considered.

### *2.3. Optimization of Parameters*

The parameter optimization scheme is conducted by creating four arrays of random $N$ numbers, $\boldsymbol{r}_i$, for randomizing $\Lambda$, $b$, $h$, and $d_{VO_2}$. The arrays consisting of random numbers are then used to produce random value of size arrays for those four parameters as: $\boldsymbol{\Lambda} = \Lambda_0 \pm \boldsymbol{r}_1\Lambda_0$, $\boldsymbol{b} = b_0 \pm \boldsymbol{r}_2 b_0$, $\boldsymbol{h} = h_0 \pm \boldsymbol{r}_3 h_0$, and $\boldsymbol{d}_{VO_2} = d_0 \pm \boldsymbol{r}_4 d_0$. Candidates with $b > \Lambda$ are rejected. The $N$ combinations of candidates are created and evaluated with the following figure-of-merit (FOM),

$$\text{FOM} = \frac{R300}{R350} = \frac{\int_{0.3\mu\text{m}}^{30\mu\text{m}} \rho_{\lambda,300} E_{\text{BB},300} d\lambda}{\int_{0.3\mu\text{m}}^{30\mu\text{m}} \rho_{\lambda,350} E_{\text{BB},350} d\lambda} \tag{3}$$

where $\rho_\lambda$ is the spectral reflectivity, and $E_{BB,T}$ is the spectral emissive power which expressed as $E_{BB,T} = \pi I_{BB,T}$, where $I_{BB,T} = (4\pi^2 c_0^2 \hbar / \lambda^5) / [\exp(2\pi\hbar c_0 / \lambda k_B T) - 1]$ is the blackbody intensity at the corresponding temperature $T$. The integrated wavelength ranges from 0.3 μm and 30 μm, which covers adequate sampling of the blackbody spectrum at atmospheric temperatures up to the Solar spectrum. The FOM is based on coating reflectivity to normalize from the opaque metal substrate, for sake of FOM comparison with other configurations on other metal substrates. The reflectivity is used to assess the coating's ability to reject surrounding radiation at higher temperature. On the other hand, for other calculations, the spectral emissivity is calculated as $\varepsilon_\lambda = 1 - \rho_\lambda$, considering that Kirchhoff's law of radiation is satisfied being the substrate is thick enough to be opaque. By fitting the parameter inputs into this summative FOM metric, further parametric stipulations can be applied to the next round of optimizations as explained in next section. Since the optimization consists of four parameters to randomize, $N = 10^4$ candidates are sufficient to search for the optimal structure. Yet optimizing more parameters could require additional samples needed, but recent machine learning techniques such as Bayesian optimization may be implemented in the future for its adaptive learning of the best parameters with reasonable convergence [39].

## 3. RESULTS AND DISCUSSION

### *3.1. Radiative Turn-Down Performance of Optimized Coating*

In this work, $BaF_2$ is chosen as the dielectric spacer and $d_{BaF_2}$ is quarter-wave thickness set constant as $\lambda_0 / 4n_{BaF_2} = 1.385$ μm. The center wavelength $\lambda_0$ is chosen as 8.2 μm, corresponding to emissive power at 80°C (hot condition). The cold condition is selected to be 30°C, a typical warm day or summer night in Texas. $BaF_2$ has a low index of approximately $n_{BaF_2} = 1.5$, which contrasts highly against $VO_2$. $BaF_2$ is preliminarily selected as the lossless dielectric due to its broad essentially zero imaginary component of the refractive index up to the Reststrahlen absorption peak near 50 μm [36]. The dielectric function of Ag is Drude-like with plasma constants from Ref. [1], and that of $BaF_2$ is obtained from Ref. [37].

First, the $d_{VO_2}$ sub-monolayer is found to have optimal thickness at 450 nm via our randomization trials. It is thinner than the anticipated quarter-wave thickness of $\lambda_0 / 4n_{VO_2} = 840$ nm, where $n_{VO_2} = 2.38$ at $\lambda = 8.2$ μm. The maximized FOM is quite broad over $N = 10^4$ samples, with no random scattering of outliers. Deviations of a few tens of nanometers from this value remains high performance. On searching for an optimal filling ratio of $VO_2$ nanowire-to-period ($\phi = w / \Lambda$), we narrowed it to the range between 0.4 to 0.6. Thus, a reasonable additional restriction can be applied to the optimization method where the width of the $VO_2$ nanowire cross-section is equal to the nanowire height. Optimizations of rectangular nanowires did not result in additional maximization of FOM. Square cross-section $VO_2$ nanowires with high length-to-width aspect ratios have been realized for nanoscale applications [23]. This round of optimizations is conducted with the aspect ratio restrictions, resulting in the best optimized structure with parameters: $\Lambda$ = 194 nm, $b$ = 114 nm, $w$ = $h$ = 80 nm, $d_{BaF_2}$ = 1.385 μm, and $d_{VO_2}$ = 0.450 μm. This design corresponds to the highest FOM = 1.47, meaning that the emitted thermal radiation at high temperature exceeds that absorbed at low temperature.

The thermal radiation turn-down mechanism can be explained starting with Fig. 2(a), where the large difference of extinction coefficient $\kappa$ between 30°C and 80°C is observed in the highlighted gray region of the spectrum. This component is close to zero at the low temperature, but well above one and increasing in magnitude in the mid to far infrared region at the high temperature. Figure 3(a) shows the averaged spectral emissivity between TM- and TE-polarized waves calculated by RCWA. The emissivity is unavoidably high at any temperature in the visible wavelengths, as there is always a bandgap transition. Meanwhile in the infrared region, the region corresponding to the extinction coefficient flip overlaps with the atmospheric transparency window (inverse of clouds on a warm day near 290 K). With the optimized design, the emissivity switches between low emissivity and high emissivity as the temperature increases. Additionally, the hemispherical emissivity ($\varepsilon^{h}$) across all polar incident angles and averaged electromagnetic wave polarizations shows small differences compared to the normal emissivity, except for the far-infrared region beyond 25 μm. This is attributed to some minor absorption due to the off-quarter-wave thickness attenuation length from a slightly lossy $BaF_2$ as it approaches its Reststrahlen band, and will be dissected in our forthcoming discussion.

The normal polar incident angle emissivity in transverse magnetic (TM) and transverse electric (TE) wave calculated in effective medium theory (EMT) are shown in Fig. 3(b), and it shows drastic differences between them. This is due to differences in effective refractive index in extraordinary and ordinary components, as expressed by [34],

$$\varepsilon_{TM} = \frac{\varepsilon_{VO_2}\varepsilon_{air}}{(1-\phi)\varepsilon_{VO_2} + \phi\varepsilon_{air}} \tag{4}$$

$$\varepsilon_{TE} = \phi\varepsilon_{VO_2} + (1-\phi)\varepsilon_{air} \tag{5}$$

where $\phi$ is the width-to-period filling ratio for gratings. The index of air is $\varepsilon_{air}$ = 1.0. The resulting effective refractive index for TM wave is calculated at around $n_{TM}$ ~1.3, and has low $\kappa_{TM}$ value for both 30°C and 80°C, which suppresses the emissivity. On the other hand, the contribution of $VO_2$ is much higher in the effective refractive index for TE waves, which $n_{TE}$ ~ 1.7 and ~3.6 at 30°C and 80°C, respectively. The $\kappa_{TE}$ value is much higher at 80°C, given by 2.0 to 4.0, and so higher emissivity is obtained. We note the near-perfect emissivity of 0.997 at 10.5 μm, attributed to the optimized design of the Fabry-Perot mechanism.

The following figures further investigate the underlying mechanisms of the turn-down coating's near-idealized performance. Figure 4 shows the emissivity calculated by RCWA as a function of both incident angle and wavelength. The results show the high turn-down performance of emissivity between cold and hot cases for both TM and TE waves. This is especially true in the neighborhood of 10 μm, as the normal and broadly oblique emissivity switches from near-zero to near-unity. The emissivity in the TM wave is slightly suppressed to due to the extinction coefficient in the effective medium extraordinary polarization being close to zero for higher temperature. On the other hand, the contrast in TE wave polarization is higher due to the switch between insulating and metallic dispersion along the nanowire. In essence, the axial dispersion along the nanowire offers the highest contrast corresponding to the TE polarization, while maintaining good transmittance overall due to the sparsity of interactions of electric fields along the nanowire cross-section. Interestingly, the high emissivity observed at far-infrared region for TM wave is not present at TE wave. EMT was also able to capture the angular polarization dependency in TM wave, where the emissive power is produced only in *x-z* plane at normal polar incident angle.

Figure 5 demonstrates the power dissipation density (PDD) plots between cold and hot states. PDD in $W/m^3$ is given by,

$$\mathrm{PDD} = \frac{1}{2}\varepsilon_0 \,\mathrm{Im}\left(\varepsilon\right)\omega\left|\mathbf{E}\right|^2 \quad (6)$$

where $\omega$ is the angular frequency corresponding to the wavelength, and $\boldsymbol{E}$ is the local electric field as function of the *x-z* plane. In Fig. 5 (a), the mechanism behind the high reflectance in the cold state is due to graded refractive index from top to bottom. With $\phi$ = 0.4, the effective refractive index of the nanowires is $n_{EMT}$ = 1.55, close to the refractive index of the dielectric, $n_{BaF_2}$ = 1.44 at 30°C and 8 μm wavelength. As the nanowire width becomes infinitesimally small the FOM plateaus close to 1.5. This makes sense as the $VO_2$ layer seeks to obtain a refractive index close to unity. The spectra calculated with an 80 nm-thick effective nanowire medium resulted the same normal emissivity, with discernable differences of off-normal spectra due to the anisotropic effective medium distribution [34]. The refractive index of the sub-monolayer $VO_2$ is $n_{VO_2} = 2.38$, which is a gradual transition to higher index before the reflecting and opaque silver substrate. The power is concentrated in a sliver at the interface between the $VO_2$ sub-monolayer and silver. Using the penetration depth definition $\delta = \lambda / 4\pi\kappa$, the attenuation depth in the silver substrate is 11 nm, almost negligible. Some absorption is seen in $VO_2$, and none through the $BaF_2$ dielectric cavity layer, as the fields are transmitted in the cold state. Since the extinction coefficient of $VO_2$ is non-zero ($\kappa = 0.08$ at 8 μm) in the transparency region shown in Fig. 2(a), the reflection is not perfect. This 5.7% $VO_2$ monolayer thickness to its penetration depth ratio is expressed correspondingly in the minimum emissivity at 8 μm shown in Fig. 3(a).

On the other hand, from Fig. 5 (b), the electromagnetic field power dissipation density (PDD) within the $VO_2$ nanowire and $VO_2$ bottom layer at the hot state temperature shows higher absorption compared to lower temperature due to its much higher $\kappa$ value of 3.3. In the hot state, the 450 nm-thick $VO_2$ sub-monolayer becomes to the metallic substrate, such that the lossless dielectric insulator becomes a Fabry-Perot resonator. This plasmonic resonance is induced within the nanowire and within the $BaF_2$ layer due to its metallic phase in vanadium dioxide layer, behaving like metal-insulator-metal (MIM) [40]. Localization of power absorption can be up to 1000 $W/m^3$ in both $VO_2$ layers. The finesse in this optimized design is due to the upper metal layer being a loosely-packed thin metal film, such that enough radiation is able to enter the dielectric cavity and constructively interfere to induce maximum absorption. Hence, the adaptive design switches from anti-reflecting coating by refractive index gradient layering to a plasmonic resonant cavity that concentrates power absorption/emission within the $VO_2$ during the coating's hot state.

Although many optical materials and coatings should perform well regardless of radiation incident angle, some angular dependence is observed at far infrared region at around 28 μm. In Figure 6, we investigate the anomalous high absorption marked by the TM wave "lobes" in Fig. 4 (a) and (b). The absorptive effect is due to a thick $BaF_2$ layer, where spectral emissivity increases with incident angle and reaches up to 1.0 at higher angles due to increasing value of $\kappa$ value and propagation length. Neither the silver substrate nor the $VO_2$ components have effects on the oblique angle emissivity. The incident far infrared light is trapped in the slightly lossy cavity and constructs multiple reflections between $VO_2$ NW and $VO_2$ sublayer. This phenomenon is only observed in TM waves at oblique incident angles, as the electric fields form surface plasmon modes at the lower corners of the dielectric $VO_2$ nanowires [41] The emissivity is further enhanced at hot state (Figs. 6 d-f) because metallic phase in $VO_2$ absorbs the light, especially the under-cavity monolayer. At steeper angles, emissive power is then concentrated within $BaF_2$ cavity, as the mechanism transitions toward cavity extinction and trapping. In practice, this angle-dependent emissivity in cold state should be avoided to improve its overall performance with radiation coming from any polarization type. With this consideration of the Reststrahlen band extinction of this particular infrared material, other dielectric materials should be considered as replacement, as discussed in the next section.

### *3.2. Performance Enhancement by Dielectric Spacer Types*

Figure 7 (a) shows a sweeping of a correspondingly quarter-wave thick layer of purely real component and wavelength-independent refractive indices from unity to Germanium ($n$ = 4.0). The $VO_2$ parameters are unchanged from the previously-found optimized configuration. Here, a higher FOM performance was found by an idealized dielectric spacer of refractive index $n$ = 2.2. This FOM can be as high as 1.86, slightly higher compared to the optimized structure with $BaF_2$ of 1.47. In this regard, we learn that other broadly infrared-transparent materials with higher index such as silicon ($n$ = 3.4) could achieve similar turn-down performance. Other low-index materials can have similar FOM as the $BaF_2$ configuration, such as potassium chloride ($n$ = 1.49) and potassium bromide ($n$ = 1.55), as long as the Reststrahlen band is far beyond the thermal infrared regime ($\lambda > 40$ μm). Ideally, medium-index materials such as zinc selenide ($n$ = 2.4) or zinc sulfide ($n$ = 2.2) can deliver FOM > 1.47. It becomes pressing to seek crystalline and interfacial compatibility between these medium-index dielectrics with $VO_2$. Difficulties may arise in more common dielectric oxides with medium-index, such as silicon dioxide or titanium dioxide, due to their Lorentz absorption in the mid- to far-infrared [42].

In the next investigation shown in Fig. 7 (b), it was found the filling ratio of the $VO_2$ NW array plays an important role in high reflectance in cold state due to index-matching between the $VO_2$ NW / air effective medium (EMT) layer and the dielectric material. The previous investigation for $BaF_2$ shows it reached a maximum at $\phi = 0.4$, but it is not true for the other dielectric materials. The highest FOM is achieved for both ZnS and the constant $n = 2.2$ index at $\phi = 0.3$. The slightly lower FOM of ZnS compared to constant $n = 2.2$ is due to its closer Reststrahlen band peak at 30.5 μm [37]. For a high-index dielectric such as Si, the best filling ratio of the $VO_2$ NW array is zero, which is unrealistic. This is due to the departure of index-gradient matching. For instance, the effective refractive index within the visible to near infrared region of EMT is $n_{EMT}$ = 1.0~2.0 for filling ratio of $\phi = 0.4$, which matches overall with the refractive index of the $BaF_2$ cavity, $n_{BaF_2}$ = 1.4~1.5. Other dielectric materials such as zinc sulfide (ZnS) and silicon (Si) may replace $BaF_2$ with an appropriate filling ratio to improve overall reflectance, but with consideration of realizable nanowire sizing. Other dielectric spacers have been used in other studies of turn-down coatings, which we will make a comparison.

### *3.3 Comparison of Radiative Turn-Down Performance*

Figures 8 (a) and (b) compares the reflectance at cold state and the emissivity at the hot state of our optimized structure ($VO_2$ nanowires / $BaF_2$ / $VO_2$ ) with a rudimentary quarter-wave (QW) layers as first-order approximation ($BaF_2$ QW / $VO_2$ QW), a Fabry-Perot resonator (thin $VO_2$ / Si) [9], and a multi-component radiative cooler system (Multilayer band-pass filter / $VO_2$ / $MgF_2$) [15]. The QW layers are presented to demonstrate its comparable high coating transmissivity as a proof of an anti-reflecting film, similar to our optimized nanowire structure in the cold state. The Planck emissive power spectrum at 80°C is also posed to predict the total emissivity between the different configurations. Table I compares the summative radiative performances. The FOM of the optimized structure is 1.47, higher than other realized structures of 0.86 and 0.6, respectively. The basic quarter-wave layers that share the same material composition as our optimized structure has a higher FOM of 1.06. The absence of the semi-metallic upper nanowire array layer produces no cavity mode interference, but shares a peak local emissivity near 8 μm. The key understanding is the absence of the upper solid $VO_2$ layer for the purpose of creating a Fabry-Perot cavity in turn reduces performance, especially the needed high reflectance in the cold state.

Thus, the $VO_2$ nanowire array is a good compromise in gaining the high turn-down, while any Fabry-Perot multilayer configuration has lower turn-down performance due to narrower band emission. We noticed some consequential differences in performance in the multi-component Fabry-Perot cooler system from our recalculations is likely due to absorption by $MgF_2$ near its Reststrahlen band near 20 μm [38], as well as differences from more modern spectroscopic surveys of $VO_2$ thin films [35, 43]. Additionally, Table 1 compares the quantities of radiative heat emitted to the surroundings. The emitted power at normal incidence is given by,

$$E_{\mathrm{emit}} = \int_{0.3\mu\mathrm{m}}^{30\mu\mathrm{m}} \varepsilon_{\lambda,T} E_{\mathrm{BB},T} d\lambda \tag{7}$$

where this quantity represents the potential to emit to a neighboring reradiating surface, such as a multilayer insulation blanket [44]. The error to these figures is no more than 10% due to the upper wavelength cutoff at room temperature and above. We also calculate the integrated hemispherical emissivity combined with absorption from atmospheric radiation, given by,

$$q''_{\mathrm{net}} = \int_{0.3\mu\mathrm{m}}^{30\mu\mathrm{m}} \varepsilon^{\mathrm{h}}_{\lambda,T} E_{\mathrm{b},T} d\lambda - \int_{0.3\mu\mathrm{m}}^{30\mu\mathrm{m}} \left(1-\tau_{\mathrm{atm},\lambda}\right) \varepsilon^{\mathrm{h}}_{\lambda,T} E_{\mathrm{BB},T_{\mathrm{amb}}} d\lambda \tag{8}$$

where $\tau_{\mathrm{atm},\lambda}$ is the atmospheric transparency shown in Fig. 2(b) [45]. The ambient temperature $T_{\mathrm{amb}}$ is assumed to be 290 K. The purpose of presenting this combined net radiative flux is to assess the respective coatings' performance exposed to diffuse cloudy or night sky. By analyzing Eq. (7), our optimized structure produces 560 W/m$^2$ in the hot state, while just 53 W/m$^2$ in the cold state. This ratio of power turn-down is 10.6:1. An even higher turn-down ratio is seen with Eq. (8), for net hemispherical heat flux with atmosphere, where the ratio is 14.7:1. The magnitudes of the hot state emissive power is not insignificant compared to the irradiative power flux from the Sun (~1400 W/m$^2$). The net radiative performance of comparable structures also show good turn-down but at predicted lower hot state magnitudes.

Although the emissive power turn-down ratio for the simpler QW layers is higher (14.3:1), the maximum emitted power is lower, therefore the reduced FOM. This FOM accounts for both the turn-down ratio as well as the emissive power difference between hot and cold state. A higher magnitude emissive power at the hot state is therefore more important, especially for compact components with small radiating surface areas. The optimized $VO_2$ NW design can also provide dense thermal radiative energy directed toward the atmosphere, almost doubling the capability for radiative rejection compared to other designs. Furthermore, we learn that the basic quarter-wave structure in fact outperforms a Fabry-Perot solid $VO_2$ layer on dielectric. This means the concept of a Fabry-Perot structure may not be best suited, compared to a simpler dual anti-reflection layer. We note the performance is bolstered by absorption within the dielectric medium at off-normal incident angles. At these emissive power magnitudes at high temperature, these coatings may be capable of passive radiative cooling under the Sun via careful tuning of spectral response in the visible wavelength.

### *3.4. Evaluation of Daytime Radiative Cooling*

This section assesses the possibility of the optimized coating as a daytime radiative cooler, and the metrics developed to quantify this capability. The Daytime Cooling Merit (DCM) is defined in Eq. (9),

$$\mathrm{DCM} = \frac{P_{\mathrm{cool},350}}{P_{\mathrm{cool},300}} = \frac{P_{\mathrm{coat},350} - P_{\mathrm{atm},350} - P_{\mathrm{sun},350} + P_{\mathrm{con},350}}{P_{\mathrm{coat},300} - P_{\mathrm{atm},300} - P_{\mathrm{sun},300} + P_{\mathrm{con},300}} \quad (9)$$

as a ratio of energy balances from the diffuse coating (coat), atmosphere (atm), the Sun (sun), and combined conduction and convective cooling (con) to an ambient temperature in atmosphere $T_{\mathrm{amb}}$ = 290 K. The coating radiative power flux in W/m$^2$ is obtained by hemispherical integration of the spectral directional emissivity with blackbody temperature function of the coating. The incoming atmospheric power flux is calculated with the diffuse atmospheric transparency spectrum seen in Fig. 3(a), and integrated with the spectral directional absorptivity at ambient temperature. We obtained the solar absorption spectrum from $I_{\mathrm{AM1.5},\lambda}$ at normal incidence spectral absorptivity, as the coating can be mechanically dialed to be sun-pointing. Lastly, heat rejection due to both convection and conduction to ambient is appropriately grouped into an overall coefficient $h_{\mathrm{c}}$, which is conservatively set as 10 W/m$^2$·K. The combined power from this component is $P_{\mathrm{con},T} = h_{\mathrm{c}}\cdot(T - T_{\mathrm{amb}})$. We note that the integration cutoff for solar irradiation at the lower wavelength of 0.3 μm gives 3% error, compared to the 9% error for the 30 μm upper limit in bodies at terrestrial temperatures.

Applying Eq. (9) onto the optimized structure results in DCM = $P_{\mathrm{cool},350}$ / $P_{\mathrm{cool},300}$ = 271 / (-615) = -0.44. This negative DCM indicates that the coating absorbs more sunlight rather than emit heat at the low temperature near ambient. However, at the high temperature which $VO_2$ becomes metallic, the coating rejects more heat than it absorbs. Since its absorptivity in the visible to near infrared region is high for any temperature as shown in Fig. 3, the coating's outgoing emissive power, $P_{\mathrm{coat},T}$ must overcome the sunlight absorption power, $P_{\mathrm{sun},T}$. In Figure 9, plotting the total radiative outgoing power with temperatures from ambient to 370 K, an equilibrium temperature is reached at around $T$ = 345 K ($P_{\mathrm{cool},T} = 0$) which is just above the phase transition temperature of vanadium dioxide at 340 K. The radiative cooling power drastically increases after passing the $VO_2$ phase transition temperature, and it can attain powers up to 470 W/m$^2$. This is similar to the net radiative cooling power at hot state in Table 1. The performance somewhat leans on the contribution of combined convection and conduction cooling power, in which a well-insulated environment around the coating may mean less cooling capability at the hot state. Therefore, the coating is able to reach self-equilibrium near the phase transition temperature of $VO_2$. In the insulating phase near ambient temperature, the coating warms up, but in the metallic phase at high temperatures, it passively cools.

This cooling capability only in the hot state exceeds that of previously designed daytime radiative cooling coatings, such as $SiO_2$ / $HfO_2$ multilayer [46], $SiO_2$-polymer hybrid metamaterial [47], and plain PDMS coatings [48]. Table 2 shows that daytime equilibrium temperature of the optimized grating is greater than ambient temperature, which does not compare with plain daytime radiative cooler materials and designs. Rather, the high cooling power just beyond the $VO_2$ transition temperature is noticeably significant compared to others. Therefore, we learn that the optimized $VO_2$ NW / $BaF_2$ / $VO_2$ is useful as a radiative cooler only when in shade and not placed under the sun. However, it is commendable for this 2 μm-thin coating to ensure that an equilibrium temperature can be obtained close to 345 K. This is not only helpful in maintaining efficiency in photovoltaic devices, but also it can be used for thermal management of components such as rechargeable batteries, fuel tanks, and engines on both terrestrial and space vehicles.

## 4. CONCLUSION

We have proposed a self-adaptive high turn-down $VO_2$ nanowire / $BaF_2$ / $VO_2$ / silver thin film of around 2 μm for passive thermal management applications. Our Monte-Carlo optimization method was utilized to obtain the optimal structure parameters for $VO_2$ components. It is compared with Fabry-Perot multilayers using $VO_2$, and a higher figure of merit (FOM) of 1.47 is obtained compared to other structures (FOM < 1). Ideally, the optimal refractive index of the dielectric spacer is $n = 2.2$, which is slightly higher than $BaF_2$ ($n = 1.55$). The electric field is reflected at the interface between $VO_2$ bottom layer and silver for lower temperature for high reflectance, but a high value of power dissipation density (PDD) is observed within the $VO_2$ in higher temperature, leading to high emissivity in infrared region. The optimized structure provides cooling power of 271 W/m$^2$ at 350 K whose power is higher compared to previously studies structures. Lastly, this work provides the idea of computational method for optimizing the photonic structures to obtain the desired optical properties. The obtained data may be utilized for machine learning (ML) technique to obtain predictive FOM with complex structural geometries and exotic materials, both dielectrics and phase transition. This structure can be extended to two-dimensional gratings to possibly further improve its figure-of-merit (FOM) as well as daytime radiative cooling merit (DCM) turn-down to an even higher value. This turn-down allows the device to self-reject the heat out of the devices such as solar cells, computer systems, engines, and functions as self-adaptive radiative cooling device.

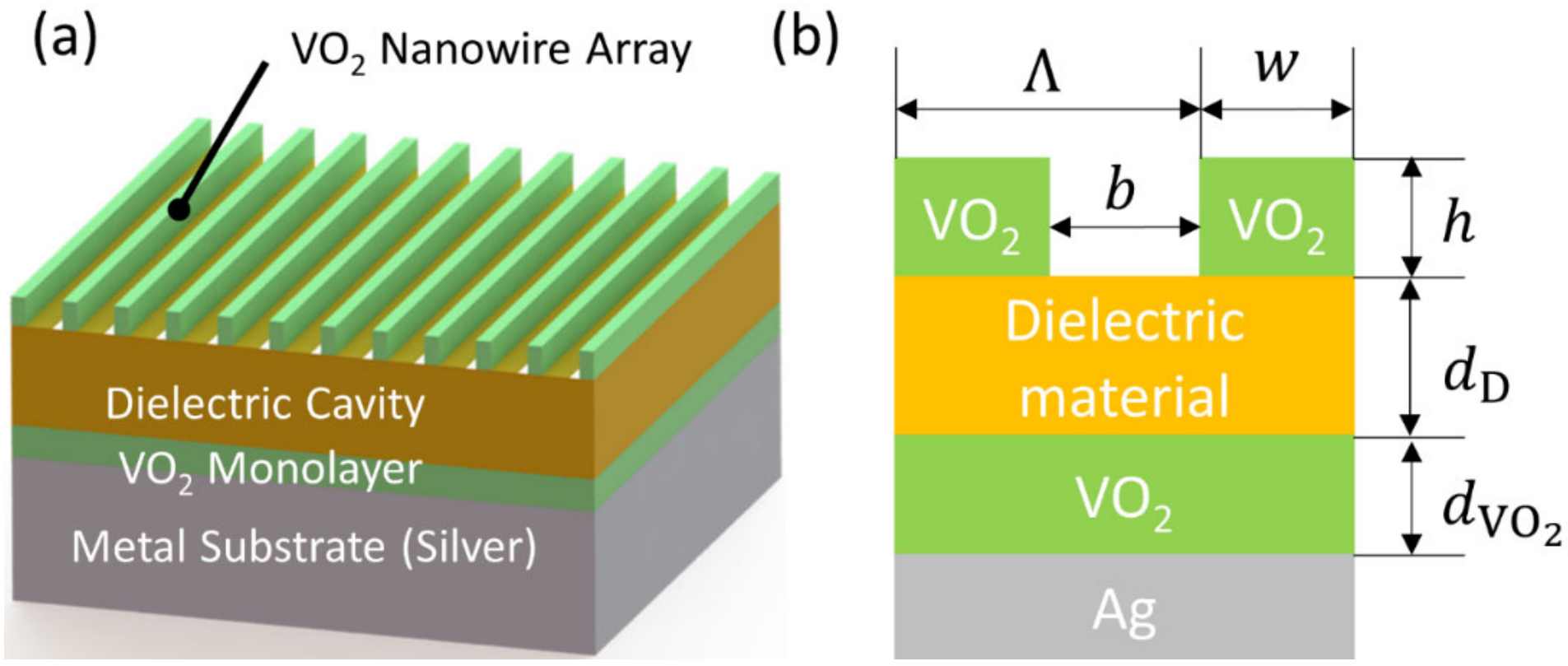

**Fig. 1.** (a) Illustration and schematic of Vanadium Dioxide ($VO_2$) nanowire / Dielectric material / $VO_2$ sub-monolayer coating on Silver (Ag) substrate. (b) Nomenclature of geometric parameters of the coating: Dielectric cavity or monolayer thicknesses $d$, nanowire grating period $\Lambda$, its width $w$, height $h$, and groove spacing $b$.

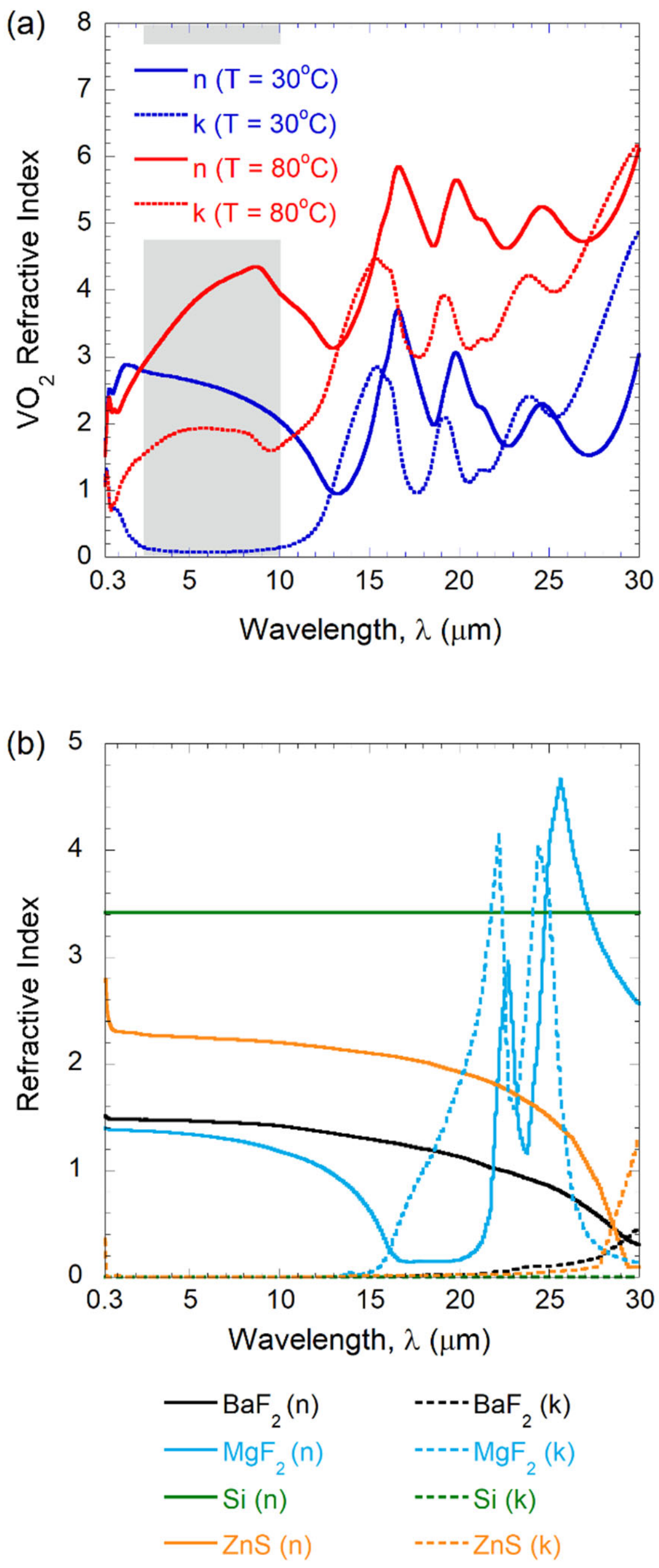

**Fig. 2.** (a) Refractive index of vanadium dioxide ($VO_2$) at $T$ = 30°C and $T$ = 80°C obtained from Ref. [35]. The highlighted region shows near-zero extinction coefficient of the refractive index ($\kappa$) at the cold state. (b) Comparison of refractive index of dielectric materials, including barium fluoride ($BaF_2$), magnesium fluoride ($MgF_2$), silicon (Si), and zinc sulfide (ZnS).

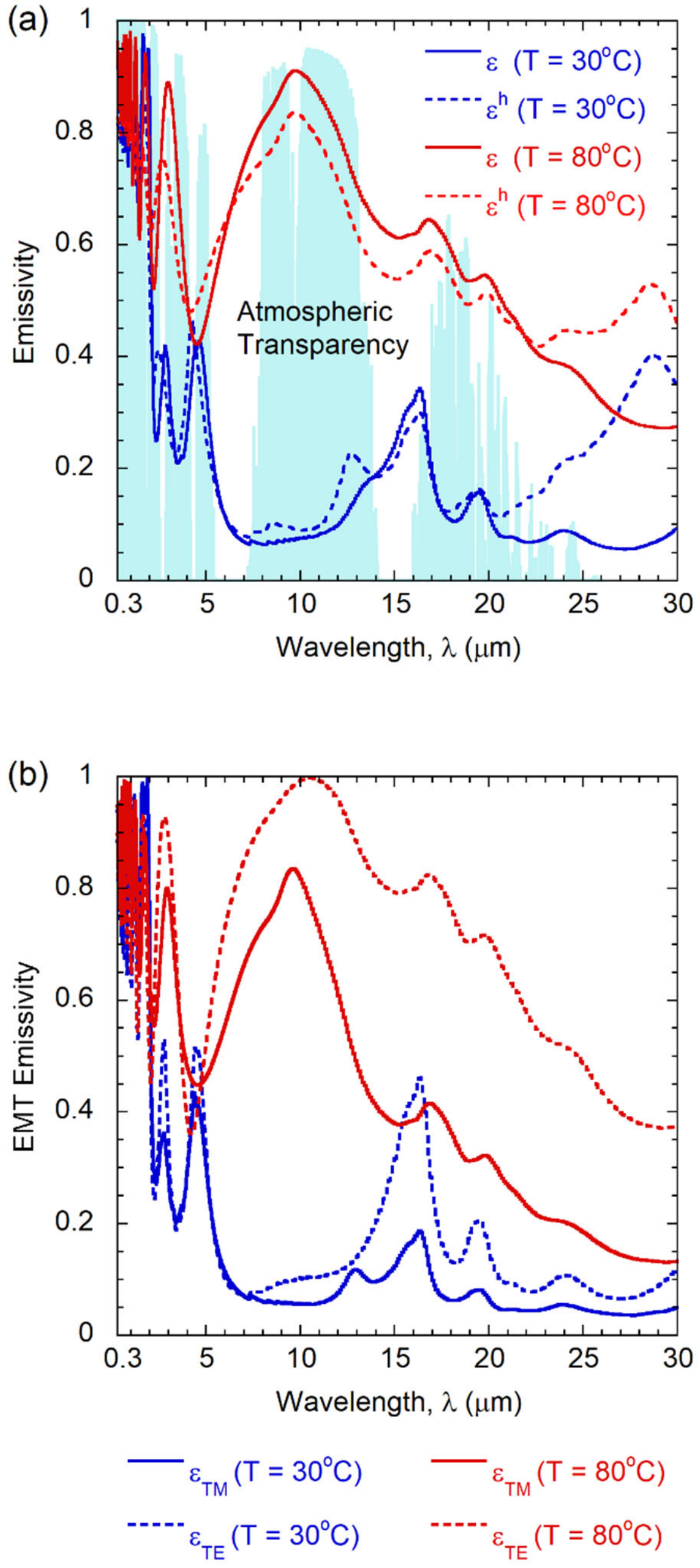

**Fig. 3.** (a) The averaged emissivity spectrum calculated with Rigorous Coupled-Wave Analysis (RCWA). The dashed lines show the hemispherical emissivity spectra. The background atmospheric transparency spectrum is shown in Ref. [45]. (b) Normal emissivity spectrum calculated with Effective Medium Theory (EMT) of optimized $VO_2$ Nanowire / $BaF_2$ / $VO_2$ coating on Ag substrate for TM (solid line) and TE wave (dashed line).

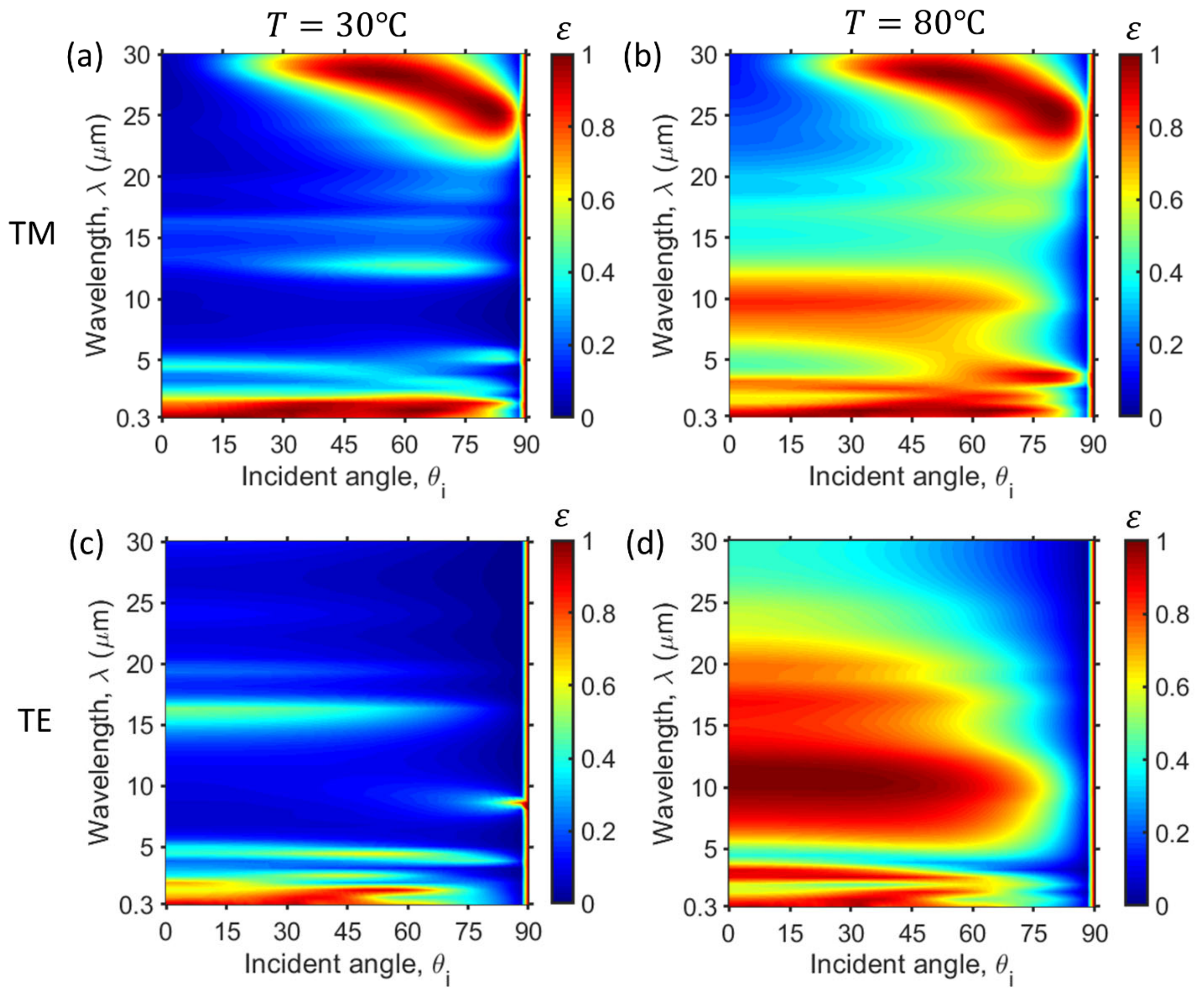

**Fig. 4.** Spectral and incident angle-dependent emissivity contours of TM wave at (a) cold and (b) hot states, and TE wave at (c) cold and (d) hot states, calculated by RCWA.

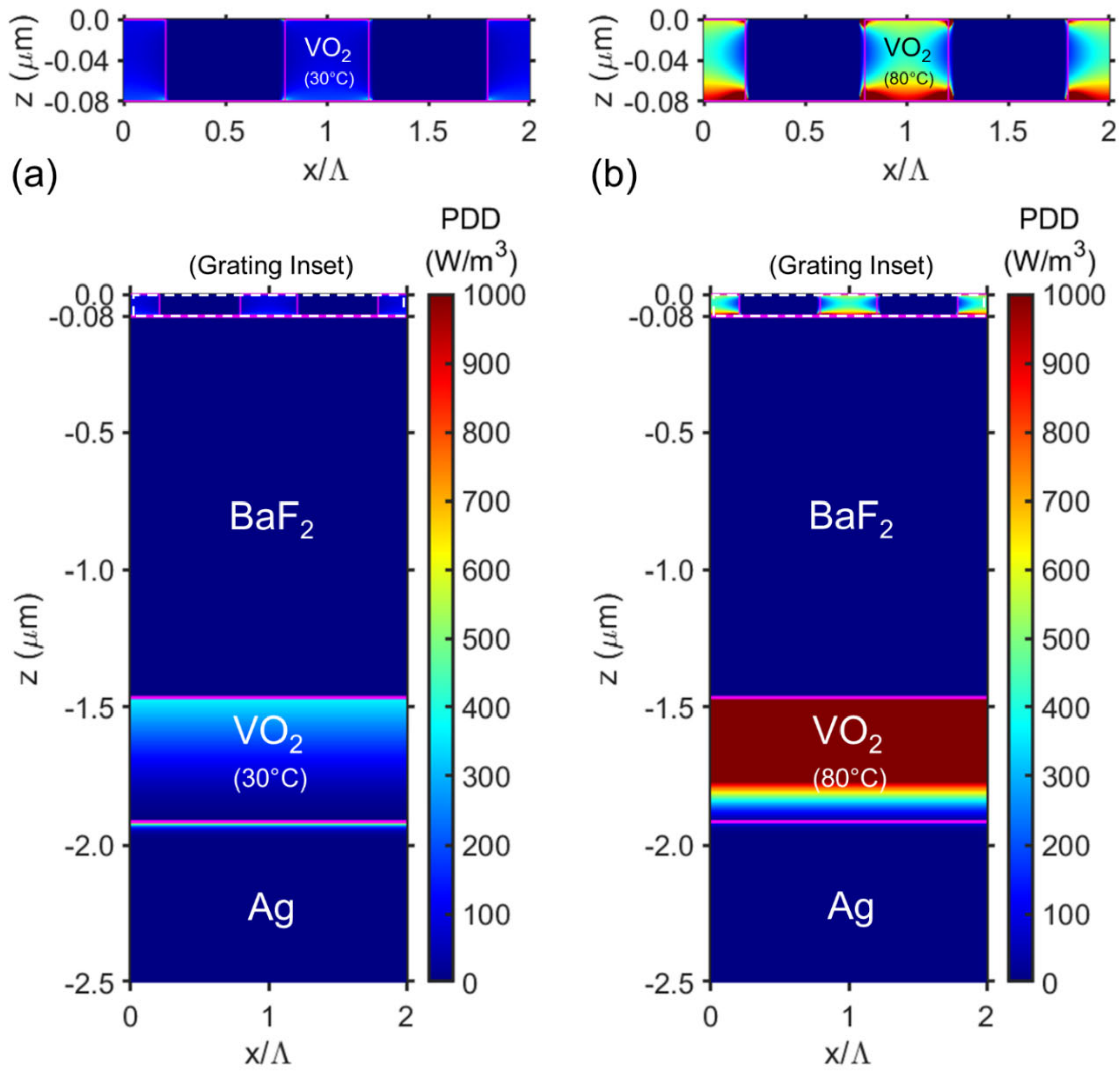

**Fig. 5.** Power dissipation density (PDD) distribution (TM wave only) of the optimized $VO_2$ NW / $BaF_2$ / $VO_2$ design for (a) cold state ($T$ = 30°C) and (b) hot state ($T$ = 80°C) calculated by RCWA. The white dashed box shows the inset figures within the $VO_2$ nanowire array region. The pink lines delineate the interfaces in the structure.

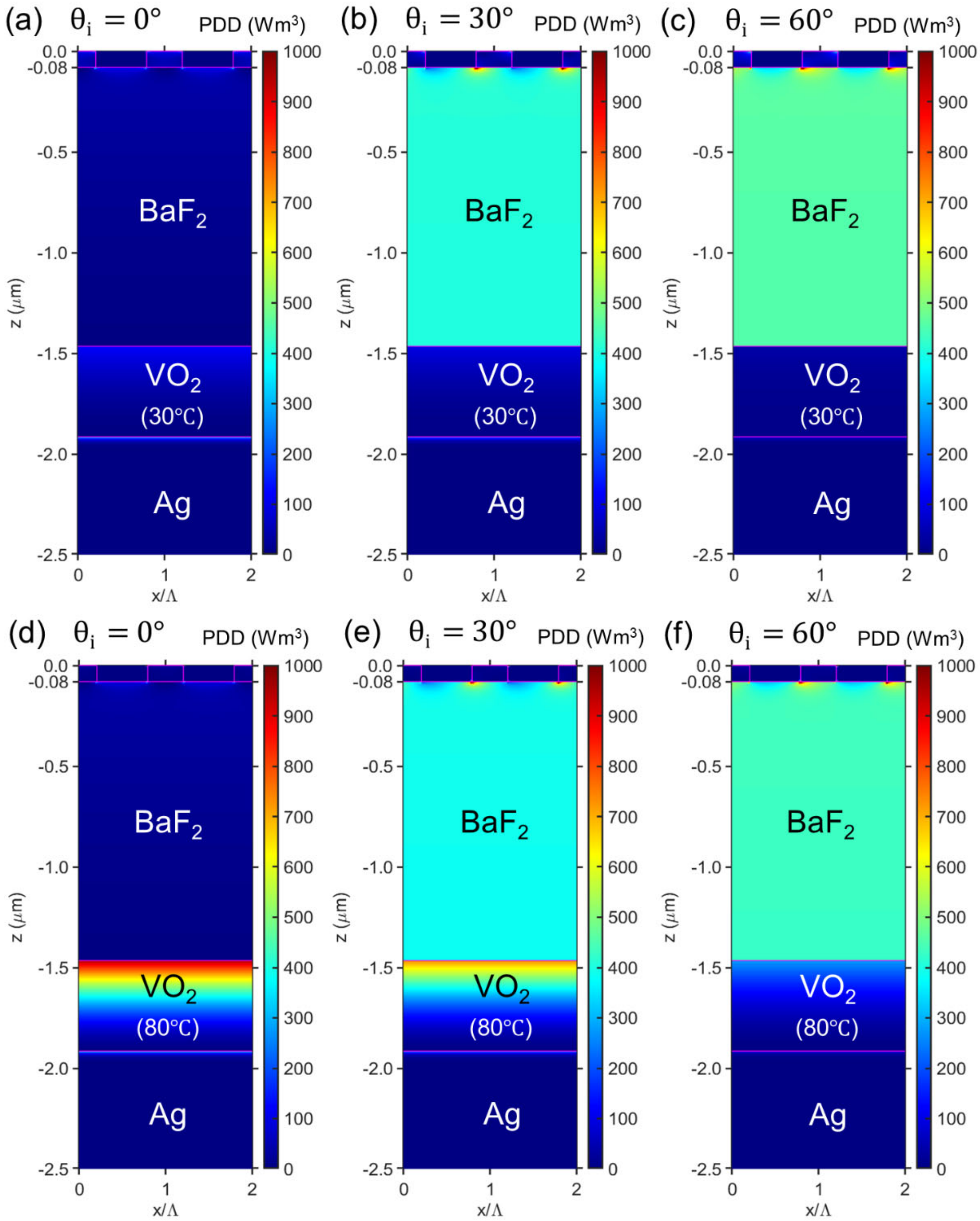

**Fig. 6.** Oblique TM wave power dissipation density (PDD) distribution of the optimized $VO_2$ NW / $BaF_2$ / $VO_2$ design at $\lambda = 28$ μm for incident angles: (a) $\theta = 0°$, (b) $\theta = 30°$, and (c) $\theta = 60°$ at cold state, and (d) $\theta = 0°$, (e) $\theta = 30°$, and (f) $\theta = 60°$ at hot state calculated by RCWA.

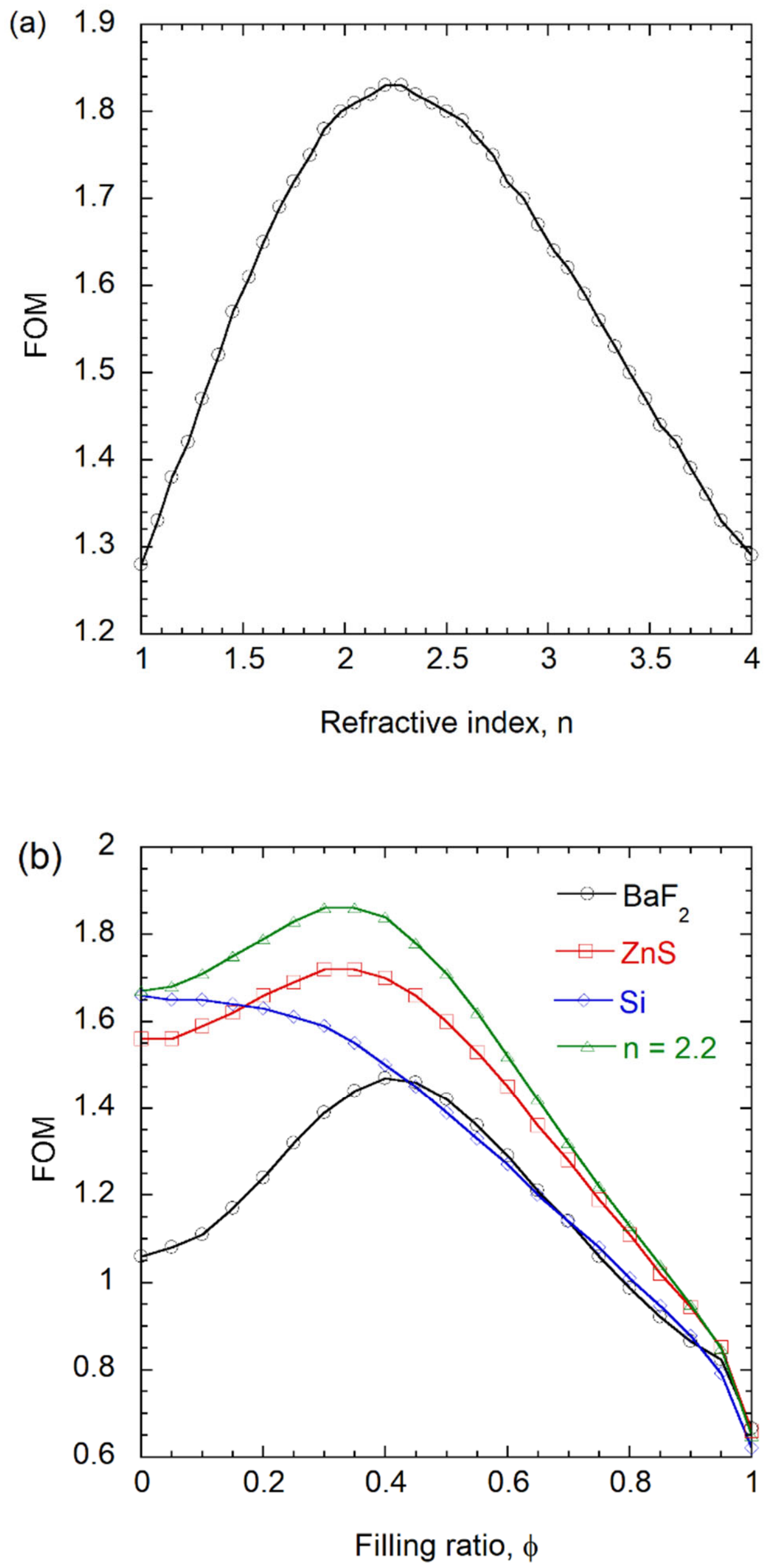

**Fig. 7.** (a) Relationship of FOM and the refractive index of the quarter-wave thick spacer. (b) The correlation of FOM and filling ratio of $VO_2$ nanowires with different dielectric materials including $BaF_2$ (black), ZnS (red), Si (blue), and constant $n = 2.2$ (green).

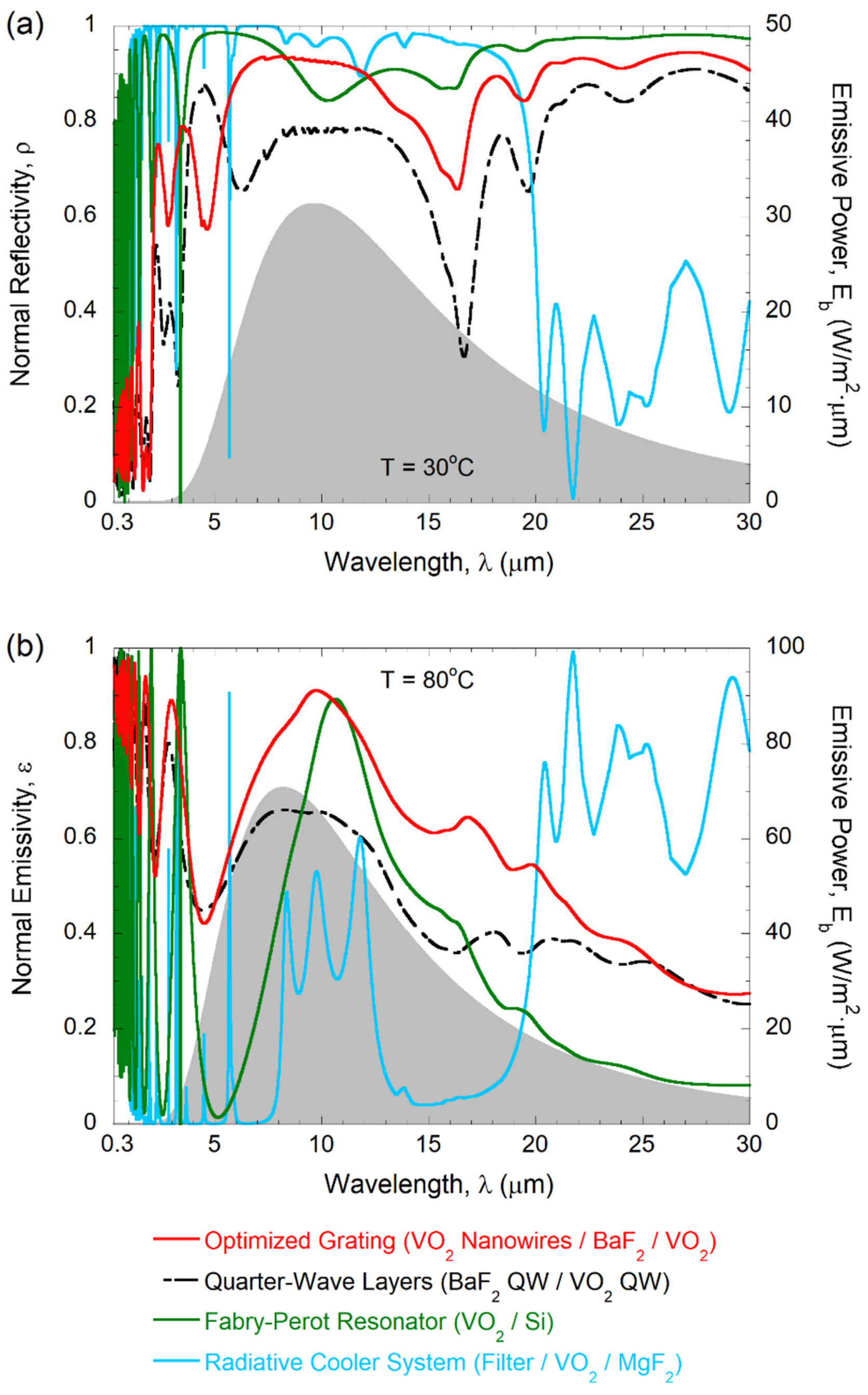

**Fig. 8.** Comparison of (a) normal reflectivity and (b) normal emissivity spectra of this optimized grating ($VO_2$ Nanowires / $BaF_2$ / $VO_2$), simple quarter-wave layers ($BaF_2$ QW / $VO_2$ QW), a Fabry-Perot resonator ($VO_2$ / Si), and a radiative cooler system (Band-pass filter / $VO_2$ / $MgF_2$). All are opaque coatings on metal substrates. The Planck blackbody emissive power function ($E_b$) at 80°C is plotted in shaded gray.

**Table 1.** Comparison of performances among similar turn-down coatings, showing the FOM defined in Eq. (3), the emitted radiation flux (W/m$^2$) from the coating ($E_{emit}$), and the net radiative heat flux between the coating and atmosphere at 290 K ( $q''_{net}$ ).

| Structure | FOM | $E_{emit}$ | | $q''_{net}$ | |
|---|---|---|---|---|---|
| | | 30°C | 80°C | 30°C | 80°C |
| Optimized Grating ($VO_2$ Nanowires / $BaF_2$ / $VO_2$) | 1.47 | 53 | 560 | 27 | 396 |
| Quarter-Wave Layers ($BaF_2$ QW / $VO_2$ QW) | 1.06 | 31 | 443 | 20 | 330 |
| Fabry-Perot Resonator ($VO_2$ /Si) | 0.86 | 34 | 359 | 18 | 265 |
| Radiative Cooler System (Filter / $VO_2$ / $MgF_2$) | 0.60 | 61 | 202 | 17 | 144 |

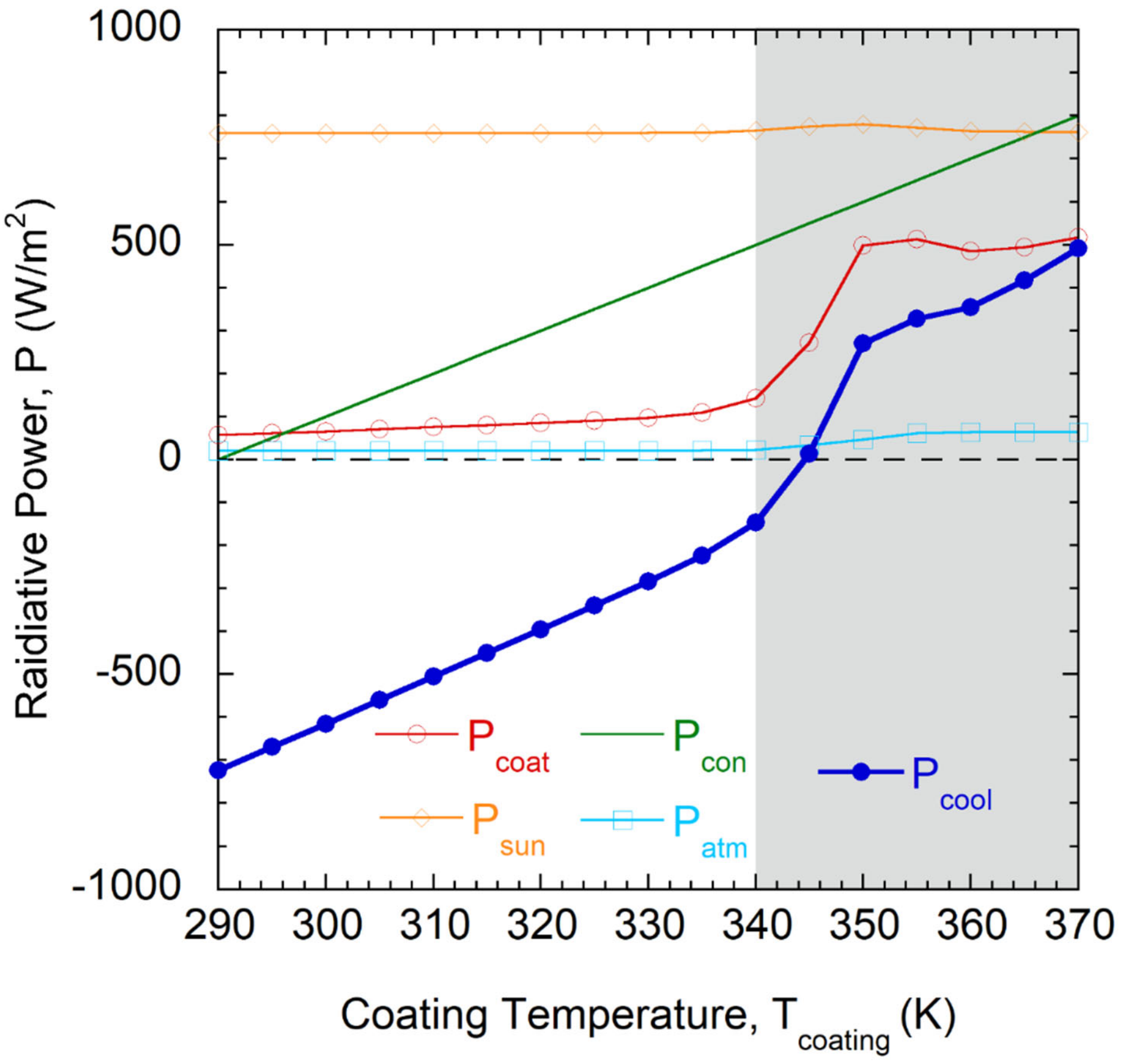

**Fig. 9.** Overall coating radiative cooling power, $P_{cool}$ and the temperature of the coating (blue line). The red, light blue, orange, and green solid lines represent $P_{coat}$, $P_{atm}$, $P_{sun}$, and $P_{con}$, respectively. The shaded area represents the metallic phase of $VO_2$.

**Table 2.** Comparison of temperature difference $T_{coating} - T_{amb}$ (K), maximum cooling power (W/m$^2$), and the Daytime Cooling Merit defined in Eq. (9), with other daytime radiative cooling coatings.

| Structure | $T_{coating}$-$T_{amb}$ (K) | Cooling power (W/m$^2$) | DCM |
|---|---|---|---|
| Optimized Grating | +46 | 271 (350 K) | -0.44 |
| $SiO_2$ / $HfO_2$ Multilayer | -4.9 | 82 | 1.86 |
| Glass-Polymer Hybrid Metamaterial | ----- | >110 | ----- |
| PDMS (300 μm) | -6 | 188 | 1.86 |